\documentclass[conference]{IEEEtran}
\IEEEoverridecommandlockouts
\usepackage{cite}
\usepackage{amsmath,amssymb,amsfonts}
\usepackage{algorithmic}
\usepackage{graphicx}
\usepackage{textcomp}
\usepackage{caption}
\usepackage{subcaption}
\usepackage{multirow}
\usepackage{comment}
\usepackage{xcolor}
\usepackage{balance}
\def\BibTeX{{\rm B\kern-.05em{\sc i\kern-.025em b}\kern-.08em
    T\kern-.1667em\lower.7ex\hbox{E}\kern-.125emX}}
\begin{document}

\title{BernoulliZip: a Compression Algorithm for Bernoulli Processes and Erdős–Rényi Graphs\\
}

\author{\IEEEauthorblockN{Amirmohammad Farzaneh,
Mihai-Alin Badiu and Justin P. Coon\IEEEauthorrefmark{1}}
\IEEEauthorblockA{\textit{Department of Engineering Science}\\
\textit{University of Oxford}\\
Oxford, United Kingdom\\
Email: \IEEEauthorrefmark{1}justin.coon@eng.ox.ac.uk}}

\maketitle

\begin{abstract}
A novel compression scheme for compressing the outcome of $n$ independent Bernoulli trials is introduced and analysed. The resulting algorithm, BernoulliZip, is a fast and near-optimal method to produce prefix codes for a Bernoulli process. BernoulliZip's main principle is to first represent the number of 1s in the sequence and then specify the sequence. The application of BernoulliZip on compressing Erdős–Rényi graphs is explored.
\end{abstract}


\section{Introduction}
Bernoulli trials \cite[Ch.~4]{dekking2005modern} are widely used in probability and statistics. They are also commonly used in numerous fields to model different events. Some of these fields include vehicular traffic models \cite{alfa1995modelling}, inventory models \cite{janssen1998r}, and disease infection \cite{kaplan1990modeling}. Surprisingly, very little work has been done on compressing finite Bernoulli processes. To the authors' knowledge, those works focus on the lossy compression of such sources \cite{wainwright2010lossy} rather than lossless compression, which is the main focus of this paper. Alternatively, arithmetic coding \cite[Ch.~13]{cover1999elements} is often used as a simple method of compression for Bernoulli processes. However, using arithmetic coding for compressing Bernoulli processes may not be feasible as the length of the sequence grows large. This is because the joint probability of Bernoulli trials decreases exponentially with the length of the sequence, and dealing with such small values can result in an error in computation. Therefore, an efficient compression technique for such processes is much needed. Moreover, it can be seen that the adjacency matrix of graphs generated using the $G(n,p)$ random graph model \cite{gilbert1959random}, also known as the Erdős–Rényi model \cite{erdHos1960evolution}, is essentially a Bernoulli process itself. Therefore, any compression algorithm used for compressing Bernoulli processes can be applied to these graphs as well. The $G(n,p)$ model is widely used in modelling the behaviour of different graphs and networks, with applications in areas such as modelling the web \cite{kumar2000stochastic}, biology \cite{10.1093/bioinformatics/bth436}, and game theory \cite{refId0}. Although there have been many methods throughout the years to compress graphs \cite{maneth2015survey}, very few have focused on compressing Erdős–Rényi graphs \cite{choi2012compression}. Consequently, a compression method for such graphs is also much needed.

In this paper, a novel approach for compressing finite Bernoulli processes is introduced. We call the algorithm BernoulliZip. It will be shown that this method is asymptotically optimal in terms of the mean length of the codewords it produces. The time complexity of BernoulliZip is also discussed, and it is shown that BernoulliZip is much faster than many existing coding methods. Additionally, the application of this method to compressing Erdős–Rényi graphs is explored, and some results from applying this method to several Bernoulli sequences and Erdős–Rényi graphs are presented.

\section{Method}

Consider the sequence \(X=\{X_1,X_2,...,X_n\}\), for which we have 
\[X_i \sim \text{Bernoulli}(p)\qquad \forall i \in \{1,..,n\}.\]
The terms Bernoulli process and Bernoulli sequence will be used to refer to such sequences. We will use $q$ throughout this text to represent $1-p$. Furthermore, consider $S$ to represent the number of 1s in $X$. In other words, we have
\begin{equation}
    S = \sum_{i=1}^{n}X_i.
\end{equation}

We will use Shannon's entropy \cite{shannon1948mathematical} to calculate the entropy of the sequence. As entropy is known to show the lossless compression bound of a source \cite[Ch.~5]{cover1999elements}, we compare the performance of BernoulliZip with it. Throughout this text,  the entropy is always calculated in base $2$. We can use conditional entropy to write
\begin{equation}
\label{entropy}
    H(X,S) =H(X) =  H(S)+H(X|S).
\end{equation}
Therefore, we divide BernoulliZip into two steps: compressing \(X|S\) and compressing \(S\).

\subsection{Compressing \(X|S\)}
\label{lexicographic}
Given $S=k$, all the possible sequences are equiprobable and we have
\begin{equation}
\label{nchoosek}
    H(X|S=k) = \log_2 {n\choose k}.
\end{equation}
Now suppose $S$ is known. To compress $X$ given $S=k$, we can simply use \(\lceil \log_2 {n\choose k}\rceil\) bits to distinguish an $n$ bit sequence from all sequences which have $k$ 1s. Notice that as the distribution of $X$ given $S$ is uniform, this representation is optimal and it is at most one bit more than the entropy. The labelling of the ${n\choose k}$ possible arrangements of the sequence can be done using a lexicographic ordering on the possible sequences. One of the possible methods for this is described below.

Suppose we have a sequence $X$ of length $n$, and consider $k$ to be the number of 1s, i.e., the Hamming weight of the sequence. If we look at $X$ as a binary number, the problem is to compute how many binary numbers with the same Hamming weight are less than $X$. In this section, we treat $X$ as a binary number with $n$ bits, with its LSB labeled as $0$ and its MSB labeled as $n-1$. For all \(1\leq i\leq  k\), $l_i$ shows the position of the $i$th $1$ from the right in the sequence. For example, $l_k$ shows the position of the most significant 1 in $X$. Consider the function \(f(X,k)\) to return the number of binary sequences less than $X$ that have the same Hamming weight as $X$. We use the notation \(X[i:j]\) to show the portion of \(X\) from position $i$ to position $j$ ($j\leq i$), inclusive. Therefore, we will have the following equation.
\begin{equation}
    f(X,k) = f(X[l_k:0],k)
\end{equation}
To calculate \(f(X[l_k:0],k)\), we divide the set of possible binary sequences into two distinct subsets: those with their MSB located at position $l_k$ and those with their MSB located at a position less than $l_k$. It can be seen that the number of sequences in the former case is \(f(X[l_{k-1}:0],k-1)\), and the latter is simply \({l_k-1\choose k}\). The following equation can be written as a result.
\begin{equation}
    f(X[l_k:0],k) = {l_k\choose k}+f(X[l_{k-1}:0],k-1)
\end{equation}
We can continue the same method, until we reach $l_1$ and the chain stops. Therefore, we will have the following equation.
\begin{equation}
    f(X,k) = \sum_{i=k}^{1}{l_i\choose i}
    \label{lexico}
\end{equation}
Notice that if at any point in (\ref{lexico}) we have \(i>l_i\), we assume \({l_i\choose i}\) to be equal to zero. This way, the binary representation of $f(X,k)$ using \(\lceil \log_2 {n\choose k}\rceil\) bits will be our codeword for $X|S$. For example, suppose we have \(X=(1101)_2\). We are interested in finding the number of binary numbers less than $X$ with a Hamming weight of 3. We can use (\ref{lexico}) and write
\[f(X,3)={3\choose3}+{2\choose 2}=2.\]
We can check that this result is in fact true, as there are only two binary numbers less than $X$ with three 1s: \((0111)_2\) and \((1011)_2\).

\subsection{Compressing $S$}

We now have to compress $S$ in an optimal manner. Notice that $S$ in fact has a binomial distribution $\text{B}(n,p)$. To achieve optimal compression, we can use Huffman coding \cite{huffman1952method} to assign codewords to the $n+1$ possible values of $S$. But instead, we will provide a much faster method. We could simply use \(\log_2 n\) bits to represent $S$. However, we will show that our method for compressing $S$ has a smaller mean length than \(\log_2 n\).

Suppose that we want to compress $S=k$. We take the following steps to find $C(k)$, the codeword for $k$.
\begin{enumerate}
    \item Set $F=0$ if $k\leq np$, or $F=1$ if $k> np$.
    \item Calculate \(t = \lfloor \log_2(|k-\lfloor np\rfloor|+1)\rfloor\), and let $T$ be its binary representation using $\lceil\log_2 \log_2 n\rceil$ bits.
    \item Calculate \(u=|k-\lfloor np\rfloor|+1-2^t\), and let $U$ be its binary representation using $t$ bits.
    \item \(C(k)=F+T+U\), where $+$ shows concatenation of the bit sequences.
\end{enumerate}
To reconstruct $k$ having $C(k)$, we identify $F$, $t$, and $u$ from the code. Then, we calculate $d = 2^{t}+u-1$. If $F=0$, we have $k =\lfloor np\rfloor-d $. Otherwise, we have $k =\lfloor np\rfloor+d $.

The intuition behind this method of compression is that we are describing the distance of $k$ from the mean value of the distribution, and assigning shorter codewords to values closer to the mean, which have a higher probability of occurrence. To this end, the binary representation of this distance is coded in the following manner. Instead of simply using \(\log_2 n\) bits, the position of the most significant bit of this number ($t$) is presented first using \(\log_2 \log_2 n\) bits. Then, $t$ bits follow to provide us with the rest of the binary number. This way, smaller distances will be paired with shorter codewords. Additionally, the existence of $F$ indicates if $k$ is lower than the mean or higher. Knowing $F$, in addition to the distance, will provide us with the exact value for $k$.

We will provide an example here to show the process of coding $k$. Assume we have $n=10$, $p=0.2$, and $k=6$. Therefore, we will have $np=2$ and as $k>np$, we will set $F=1$. We then have \(t=\lfloor \log_2(|6-2|+1)\rfloor=2\), \(u=|6-2|+1-2^2=1\), and \(U=(01)_2\). As \(\lceil\log_2 \log_2 10\rceil=2\), we will have \(T=(10)_2\). Finally, we will have \(C(6)=11001\).

Based on the discussions, the codeword for $X$ is the code for its number of 1s using the described method, followed by the lexicographic order of $X$ among all sequences with the same Hamming weight. Notice that BernoulliZip always produces a prefix code. After observing the initial $1+\lceil\log_2 \log_2 n\rceil$ bits of the code, $t$ will be known. The following $t$ bits will help us identify $k$, and then \(\lceil \log_2 {n\choose k}\rceil\) bits will follow to determine the lexicographic ordering of the sequence.

\section{Mean length}
In this section, we will calculate an upper bound for the mean length of the proposed compression scheme. The codeword consists of two parts: $S$ and the index of $X$ given $S$. If we show the length of the codeword for $X$ with $L$ and the length of $C(k)$ with $L_k$, we have the following equation.
\begin{equation}
    \label{length}
    \mathbb{E}(L) = \sum_{k=0}^{n}{n\choose k}p^kq^{(n-k)}(L_k+\lceil\log_2{n\choose k}\rceil)
\end{equation}
As we already know that the encoding of $X$ given $S$ is optimal, we are interested in calculating the mean length of our encoding method for $S$. Firstly, notice that based on the described algorithm, we have the following equation for the length of $C(k)$.
\begin{equation}
\label{lck}
    L_k = 1+\lceil\log_2 \log_2 n\rceil+\lfloor\log_2(|k-\lfloor np\rfloor|+1)\rfloor
\end{equation}
For large values of $n$, we can ignore the ceiling and floor functions in (\ref{lck}) and write
\begin{equation}
    \label{ineq 1}
        \mathbb{E}[L_k] = 1+\log_2 \log_2 n+\mathbb{E}[\log_2(|k- np|+1)].
\end{equation}
As the logarithm is a concave function, we can use Jensen's inequality \cite[Th.~2.6.2]{cover1999elements} and write
\begin{equation}
    \label{Jensen}
    \begin{split}
        \mathbb{E}[\log_2(|k-np|+1)]&\leq\log_2(\mathbb{E}[|k-np|+1])\\
        &=\log_2(\mathbb{E}[|k-np|]+1)
    \end{split}
\end{equation}
As $k$ comes from a $\text{B}(n,p)$ distribution, \(\mathbb{E}[|k-np|]\) is actually the mean absolute deviation (MAD) of the binomial distribution for which we have the following inequality \cite{berend2013sharp}.
\begin{equation}
    \label{mad}
    \mathbb{E}[|k-np|]\leq \sqrt{npq}
\end{equation}
By inserting (\ref{mad}) into (\ref{Jensen}), we will have
\begin{equation}
    \label{temp1}
    \mathbb{E}[\log_2(|k-np|+1)]\leq \log_2(\sqrt{npq}+1).
\end{equation}
Ultimately, (\ref{lck}) and (\ref{temp1}) will provide us with the following upper bound on the mean length of our code for the binomial distribution.
\begin{equation}
    \label{binomial}
    \begin{split}
        \mathbb{E}[L_k]&\leq \log_2(\sqrt{npq}+1)+\log_2 \log_2 n+1\\
        &\leq \log_2(\sqrt{npq})+\log_2 \log_2 n+2
    \end{split}
\end{equation}

We can now calculate an upper bound for the mean code length of BernoulliZip. The entropy of the original sequence is in fact the entropy of $n$ independent Bernoulli trials and equals $nh(p)$, where \(h(p)\) is the entropy of a single Bernoulli trial and equals \(-p\log_2p-(1-p)\log_2(1-p)\). The difference between BernoulliZip's mean code length and the entropy of the sequence is due to the difference between $\mathbb{E}[L_k]$ and the entropy of $B(n,p)$. Additionally, there is a difference of at most 1 bit between BernoulliZip's compression of \(X|S\) and \(H(X|S)\). As the entropy of $B(n,p)$ is known to be \(\frac{1}{2}\log_2(2\pi enpq)+O(\frac{1}{n})\) \cite{knessl1998integral}, we will have the following upper bound on BernoulliZip's mean code length for large $n$s.
\begin{equation}
    \label{code upper bound}
    \begin{split}
        \mathbb{E}(L)&\leq nh(p)+\mathbb{E}[L_k]-\frac{1}{2}\log_2(2\pi enpq)+1\\
        &=nh(p)+\log_2 \log_2 n+\log_2\sqrt{\frac{1}{2\pi e}}+3
    \end{split}
\end{equation}

It can be seen in (\ref{code upper bound}) that the mean code length of BernoulliZip is at most of order \(\log_2\log_2 n\) bits longer than the entropy of the sequence. Therefore, the mean codeword length of BernoulliZip is asymptotically optimal and we have
\begin{equation}
\label{asymp}
    \mathbb{E}(L)\sim nh(p).
\end{equation}

\section{Time complexity analysis}
\label{time complexity}

In this section, we will calculate the time complexity of BernoulliZip, when run on a sequence of length $n$. The algorithm consists of different parts for which we will calculate the time complexity individually.
\begin{itemize}
    \item\textbf{Finding the number of 1s in the sequence:} Finding the Hamming weight of a given sequence is the first step we take when compressing the sequence. This is a basic summation of the sequence, and is of $O(n)$ complexity.
    \item\textbf{Lexicographic ordering:} Given the sequence and its number of 1s, we must calculate its lexicographic order. One possible method was described in section \ref{lexicographic}, whose time complexity will provide us with an upper bound on the time complexity of this step. It can be seen that this order can be found using a summation on the sequence according to (\ref{lexico}), and is at most of $O(n)$ complexity. However, this is only if we assume the values of the binomial coefficients as given, so that there is no need for calculating them in each step of the summation. This is a valid assumption if $n$ is not very large.
    \item\textbf{Calculating $t$:} Calculating $t$ requires the calculation of the floor of a logarithm, with its maximum input being equal to $n$.  This can be done using at most $\log_2 n$ multiplications by 2. Therefore, this step is at most of $O(\log n)$ complexity.
    \item\textbf{Calculating $u$:} To calculate $u$, a power function needs to be calculated, with the exponent being at most equal to $\log_2 n$. Simple implementations of the power function will do this using at most \(\log_2 n\) multiplications. Consequently, this step is also of $O(\log n)$ time.
    \item\textbf{Finding the binary representation of $t$ and $u$:} $t$ is a number less than $\log_2 n$, and $u$ is a number less than $n$. Therefore, representing each of them in binary can be done using at most \(\log_2 n\) divisions, and is therefore of at most $O(\log n)$ complexity.
\end{itemize}
It can be seen that all steps of the compression process are of $O(n)$ complexity or less. Therefore, BernoulliZip has a time complexity of at most $O(n)$, with $n$ being the length of the sequence.

We compare the time complexity of BernoulliZip with two coding algorithms that have an optimal mean code length: Huffman coding and Shannon-Fano-Elias coding \cite[Ch.~5]{cover1999elements}. For a source with $A$ alphabets, Huffman coding has $A$ steps. In each step, the two alphabets with the lowest probabilities are found and combined, which is of $O(\log A)$ complexity. Therefore, Huffman coding is of $O(A\log A)$ complexity. To code a symbol $s$ from a source with $A$ alphabets, Shannon-Fano-Elias coding performs a summation on the probability of all symbols less than $s$. This makes Shannon-Fano-Elias coding of at least $O(A)$ complexity. Notice that for a Bernoulli process of length $n$, we have $A=2^n$. Table \ref{tab3} compares BernoulliZip with these two methods in terms of complexity and mean length. It can be observed that in addition to BernoulliZip having an asymptotically optimal mean length, it has a much lower complexity than the two other algorithms.

\begin{table}[h!]
    \caption{Comparison of different algorithms for compressing a Bernoulli process of length $n$ and parameter $p$}
    \centering
    \begin{center}
    \resizebox{\columnwidth}{!}{
    \begin{tabular}{|c|c|c|}
    \hline
        \textbf{Algorithm} &  \textbf{Time Complexity} & \textbf{Mean Length}\\
        \hline
         BernoulliZip& $O(n)$& \(nh(p)+\log_2 \log_2 n\)\\
         Huffman & $O(n2^{n})$ & \(nh(p)\)\\
         Shannon-Fano-Elias & $O(2^{n})$ & \(nh(p)+2\)\\
         \hline
    \end{tabular}
    }
    \end{center}
    \label{tab3}
\end{table}

\section{Application to Erdős–Rényi graphs}

An Erdős–Rényi graph with $v$ nodes can be represented by a binary sequence of length $v\choose 2$, where each bit represents the existence of each edge. Each edge exists independently with probability $p$. Therefore, the resulting sequence is essentially a Bernoulli process and its entropy equals \({v\choose 2}h(p)\). Consequently, we can use BernoulliZip to compress these graphs. There are two ways to apply BernoulliZip to an Erdős–Rényi graph with $v$ vertices.

\begin{enumerate}
    \item \textbf{Direct method:} In this method, we apply BernoulliZip to the whole graph as a Bernoulli sequence of length \(v\choose2\). Based on (\ref{code upper bound}), the average generated code length in this case will be roughly \(\log_2 \log_2 {v\choose 2}\) bits longer than the entropy of the process. By inserting $v\choose 2$ as the value of $n$ in (\ref{code upper bound}), we can see that the mean code length of the direct method for graphs is also asymptotically optimal and we have \(\mathbb{E}(L)\sim {v\choose 2}h(p)\). However, as $v$ grows large, the time complexity of the direct method increases with $v^2$.
    
    \item \textbf{Block method:} In this method, we fix a block length $n$, divide the graph's edge sequence into blocks of length $n$, and then apply the method to each block separately. In this case, the mean code length will be approximately $\frac{{v\choose2}}{n}\log_2 \log_2 n$ bits longer than the entropy of the sequence. Even though the mean code length in this method is larger than the direct method, this method gives us the ability to compute the codewords of different blocks in parallel, which results in faster execution of the algorithm.
\end{enumerate}

To choose between these two methods and the value for the block size $n$, we must take into account the size of the graph, the complexity of implementing the algorithm, and the amount of compression needed.

\section{Results}

\begin{figure}[h!]
\caption{Entropy and mean compressed length of Bernoulli processes with length $50$ as a function of $p$.}
    \centering
    \includegraphics[width = \columnwidth]{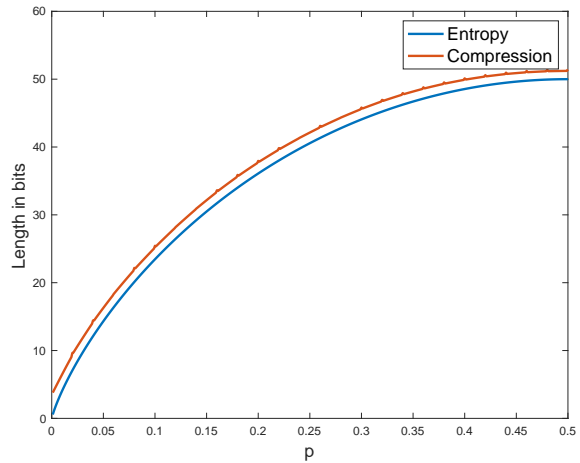}
    \label{fig1}
\end{figure}

\begin{figure}[h!]
\caption{Entropy and mean compressed length of Bernoulli processes with $p=0.1$ as a function of the length of the Bernoulli sequence, $n$.}
    \centering
    \includegraphics[width = \columnwidth]{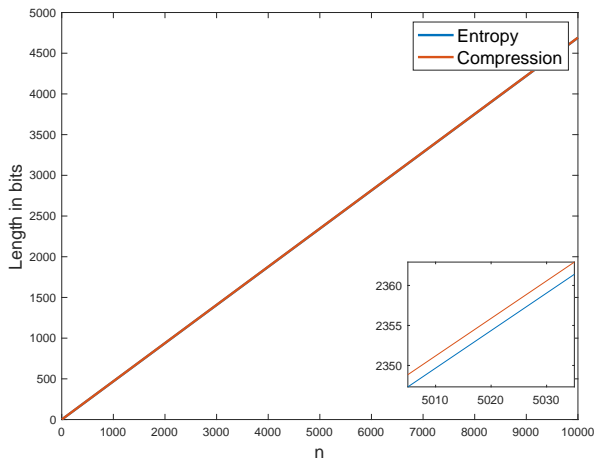}
    \label{fig2}
\end{figure}

In this section, we present the results obtained from simulating BernoulliZip in MATLAB. Table \ref{tab1} and Table \ref{tab2} show the results of applying BernoulliZip to a number of Bernoulli processes and Erdős–Rényi graphs using the direct method and block method, respectively. For each model, a large number of instances were created and then compressed. The mean length of each model was calculated as the average length of the compressed codewords. The standard deviation of the compressed code lengths are also reported. It can be seen that the mean compressed length is very close to the entropy of the sequence for most of the cases. Additionally, the direct method clearly has a better performance than the block method. Figure \ref{fig1} illustrates the mean length of BernoulliZip for Bernoulli sequences of length $50$ as a function of the Bernoulli parameter $p$, and compares it with the entropy of the sequence. Figure \ref{fig1} indicates the mean code lengths of BernoulliZip to be near the entropy of the sequence for different values of $p$. Figure \ref{fig2} shows the change in the mean length of BernoulliZip as $n$ grows large, for a constant value of $p$. It can be seen that there is only a slight difference between the entropy of the sequence and the mean length of BernoulliZip. A zoomed section of the plot is also illustrated to show the gap, as it can not be seen in the original plot. All in all, the simulation results are indicating the near-optimal performance of BernoulliZip in terms of its mean code length.

\begin{table}[h!]
    \caption{Results of running BernoulliZip on a number of Bernoulli processes and Erdős–Rényi graphs (direct method)}
    \centering
    \begin{center}
    \resizebox{\columnwidth}{!}{
    \begin{tabular}{ |c||c|c|c|}
     \hline
     \multirow{2}{*}{\textbf{Sample}}    & \multirow{2}{*}{\textbf{Entropy}} & \multicolumn{2}{c|}{\textbf{Direct method}}\\
     \cline{3-4}
     &&Mean length & Standard deviation\\
     \hline
     \(\text{Bernoulli}(50,0.1)\)  & 23.4498 b & 25.6430 b &7.7285\\
     \hline
      \(\text{Bernoulli}(50,0.01)\)  & 4.0397 b & 6.9080 b&2.9597\\
      \hline
      \(\text{Bernoulli}(20,0.2)\)  & 14.4386 b & 17.1340 b&4.4790\\
      \hline
      \(G(5,0.1)\)  & 4.6900 b & 6.8081 b&2.4765\\
      \hline
      \(G(8,0.2)\)  & 20.2140 b & 22.6220 b&4.2082\\
      \hline
      \(G(10,0.1)\) & 21.1048 b & 23.6380 b&6.1193\\
      \hline
    \end{tabular}
    }
    \end{center}
    \label{tab1}
\end{table}

\begin{table}[h!]
    \caption{Results of running BernoulliZip on a number of Bernoulli processes and Erdős–Rényi graphs (block method)}
    \centering
    \begin{center}
    \resizebox{\columnwidth}{!}{
    \begin{tabular}{ |c||c|c|c|c|}
     \hline
     \multirow{2}{*}{\textbf{Sample}}    & \multirow{2}{*}{\textbf{Entropy}} & \multicolumn{3}{c|}{\textbf{Block method}}\\
     \cline{3-5}
     &&Block length & Mean length & Standard deviation\\
     \hline
     \(\text{Bernoulli}(200,0.2)\)  & 144.3856 b& 5 & 234.8850 b & 9.7552\\
     \hline
      \(\text{Bernoulli}(1000,0.01)\)  & 80.7931 b& 50 & 139.8860 b & 18.3887\\
      \hline
      \(G(20,0.05)\)  & 54.4154 b& 10 & 92.5930 b & 10.5713\\
      \hline
      \(G(100,0.01)\)  & 399.9260 b & 25 & 1039.2030 b & 34.8881\\
      \hline
    \end{tabular}
    }
    \end{center}
    \label{tab2}
\end{table}

\section{Conclusion}

BernoulliZip has been presented as a novel approach to compress the outcome of a finite Bernoulli process. This efficient method produces codes with a mean length that is close enough to the entropy of the sequence of Bernoulli trials. At the same time, BernoulliZip is fast and easy to implement. In other words, BernoulliZip's advantage to previous methods is its asymptotically optimal mean code length, combined with its low computational complexity. Additionally, the direct and block methods were introduced as two different approaches for applying BernoulliZip to graphs that are created using the $G(n,p)$ model. The results from simulating BernoulliZip exhibited its great performance on both raw Bernoulli sequences and Erdős–Rényi graphs.

\section*{Acknowledgment}
This work was supported by EPSRC grant number EP/T02612X/1.

\balance

\end{document}